*Annual Review of Earth and Planetary Sciences*

# The Role of Giant Impacts in Planet Formation

## Travis S.J. Gabriel[1,*] and Saverio Cambioni[2,*]

[1]Astrogeology Science Center, U.S. Geological Survey, Flagstaff, AZ, USA;
email: tgabriel@usgs.gov

[2]Department of Earth, Atmospheric and Planetary Sciences, Massachusetts Institute of
Technology, Cambridge, MA, USA; email: cambioni@mit.edu

## Abstract

Planets are expected to conclude their growth through a series of giant impacts: energetic, global events that significantly alter planetary composition and evolution. Computer models and theory have elucidated the diverse outcomes of giant impacts in detail, improving our ability to interpret collision conditions from observations of their remnants. However, many open questions remain, as even the formation of the Moon—a widely suspected giant-impact product for which we have the most information—is still debated. We review giant-impact theory, the diverse nature of giant-impact outcomes, and the governing physical processes. We discuss the importance of computer simulations, informed by experiments, for accurately modeling the impact process. Finally, we outline how the application of probability theory and computational advancements can assist in inferring collision histories from observations, and we identify promising opportunities for advancing giant-impact theory in the future.

■ Giant impacts exhibit diverse possible outcomes leading to changes in planetary mass, composition, and thermal history depending on the conditions.

■ Improvements to computer simulation methodologies and new laboratory experiments provide critical insights into the detailed outcomes of giant impacts.

■ When colliding planets are similar in size, they can merge or escape one another with roughly equal probability, but with different effects on their resulting masses, densities, and orbits.

■ Different sequences of giant impacts can produce similar planets, encouraging the use of probability theory to evaluate distinct formation hypothesis.

**Planetary embryo:**
solid planetary body that is the precursor of a planet (sometimes synonymous with protoplanet); thought to be Moon sized or larger

## 1. INTRODUCTION

The last stage of planet formation is thought to have been marked by a period of mutual collisions, called giant impacts (see the sidebar titled What Makes an Impact Giant?). For the Solar System, these collisions have been proposed to involve Moon- to Mars-sized planetary embryos, eventually resulting in a few terrestrial planets remaining in widely spaced, stable orbits (**Figure 1**). Because of the similar size of the colliding bodies and the large range of their expected impact velocities (up to tens of kilometers per second), giant impacts are global-scale, high-energy events that

---

### WHAT MAKES AN IMPACT GIANT?

When impactors are small, their imprints generally resemble a bowl-shaped divot or a wide, flat impression, deemed crater-forming or basin-forming events, qualifiers that depend on the outcome. When the impactor and target are comparable, such as the Moon-forming impact or collisions in the asteroid belt, the common monikers are pairwise accretion and similar-sized collisions; aside from gas giants, there is little distinction between the terms similar size and similar mass. When the similarly sized colliding bodies are large and their interaction is dominated by gravity, the term giant impacts is often employed. The terms similar-sized collisions and giant impacts are not based on outcome as the results vary considerably under expected impact parameters. The exact transition from cratering to giant impacts remains unclear.

---

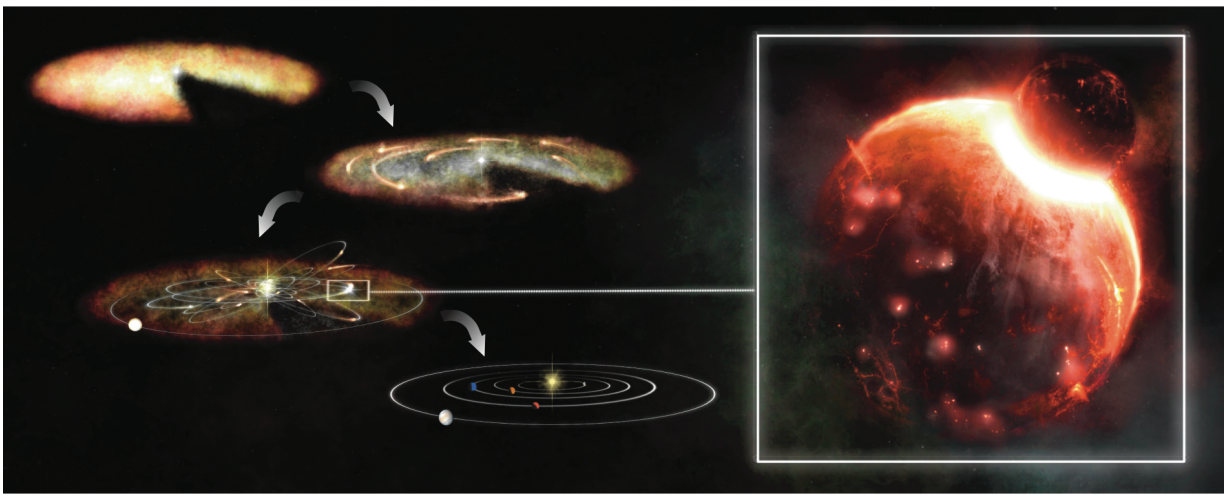

**Figure 1**

A simplified view of the classical model for terrestrial planet formation (not to scale). From top to bottom: The central star is surrounded by nebular gas and dust where early solids form. In the next stage, nebular gas begins to dissipate over ~2–3 million years (Williams & Cieza 2011), and mass accumulates into fewer and larger bodies, forming planetary embryos. Next, the orbits of the planetary embryos cross each other and lead to giant impacts such as that illustrated in the inset. After tens to hundreds of millions of years of giant impacts, the terrestrial planets achieve a stable architecture (for a review, see Raymond & Morbidelli 2022) (see Section 3). Figure adapted from Levin (1972); image courtesy of Andrew Gonzalez.

can alter planetary compositions and atmospheres that planets inherited from nebular processes. For example, primordial volatiles or atmospheres may be driven off by a giant impact, or mantle materials can be stripped and ejected entirely. Magma oceans are also often a consequence of giant impacts. Heat production in large fractions of the planets can influence geodynamic processes and promote the outgassing of volatiles previously bound within planetary interiors. The consequences of these effects on planetary habitability are not yet entirely understood.

Analytical work and computer simulations have revealed that giant-impact outcomes can vary considerably depending on the impact conditions. In some scenarios, the colliding bodies merge into a larger body, with minor amounts of material ejected in the process. For off-axis scenarios at sufficient velocity, the bodies can graze each other and escape along deflected paths. Although expected to be rare in Solar System formation, highly erosive outcomes are also possible, resulting in destruction of the target and/or impactor depending on preimpact conditions. Crust and mantle materials are ejected to varying degrees in all scenarios causing changes in composition. As an important by-product, ejected debris can collide with other planets or contribute to the population of small bodies such as moons and asteroids.

The diversity of giant-impact outcomes may be the cause of the wide array of planetary bulk compositions observed in the Solar System and exoplanetary data. However, addressing this hypothesis is challenging because different impact conditions can produce similar outcomes. For example, several models of the Moon-forming giant impact can reproduce the bulk properties of the Earth-Moon system with similar fidelity. Furthermore, different planetary systems can form from the same initial conditions due in part to the chaotic nature of gravitational interactions. Understanding the diversity of giant-impact outcomes and incorporating them into models of planet formation are critical steps to address competing hypotheses for the formation of Solar System and exoplanetary bodies.

## 1.1. Basics of Giant-Impact Theory

Study of the giant-impact phenomenon started in the mid-twentieth century as part of the larger debate on nebular theory and the coagulation of early Solar System solids (e.g., Schmidt 1958). Analytical studies of the evolution of protoplanets before and after nebular gas dissipation showed that collisions played a vital role in both stages. Without the presence of gas, two bodies generally collide at a velocity ($v_{coll}$) larger than or equal to their mutual escape velocity,

$$v_{esc} = \sqrt{\frac{2\mathcal{G}(M_T + M_I)}{(R_T + R_I)}}, \qquad 1.$$

where $\mathcal{G}$ is Newton's gravitational constant, $M$ is mass, and subscripts T and I indicate the target and impactor, respectively (e.g., Gurevich & Lebedinsky 1950). An impact velocity of $v_{coll} = v_{esc}$ is the result of gravitational forces between two bodies accelerating toward one another from an infinite distance with no initial relative velocity ($v_\infty = 0$). It is derived by equating the impact kinetic energy,

$$K = \frac{1}{2}\frac{M_T M_I}{M_T + M_I}v_{coll}^2, \qquad 2.$$

to the orbital potential energy of the bodies at contact,

$$U_{\mathcal{G},contact} = \mathcal{G}\frac{M_T M_I}{(R_T + R_I)}, \qquad 3.$$

**WHAT IS THE MOST PROBABLE COLLISION ANGLE ($\theta_{coll}$)?**

The differential probability that a point-mass impactor will hit a massless spherical target at a distance $x = R_T \sin\theta_{coll}$ from its center is $dP = 2\pi x dx = 2\pi R_T^2 \sin 2\theta_{coll} \, d\theta_{coll}$ with a maximum at $\theta_{coll} = 45°$ (Gilbert 1893). For an undeformable target with mass, the gravitational cross section $R$, which represents the maximum distance of an impactor's undeflected path from the target that would result in impact, exceeds $R_T$. From conservation of angular momentum, $x = R_T v_{coll}/v_\infty \sin\theta_{coll}$ yielding $dP = 2\pi x dx = 2\pi R_T^2 (v_{coll}/v_\infty)^2 \sin 2\theta_{coll} \, d\theta_{coll}$, and peaking at $\theta_{coll} = 45°$ (Shoemaker 1962). The $dP(\theta_{coll}) \propto \sin(2\theta_{coll})$ proportionality is expected to hold for giant impacts, although tidal distortions may complicate matters (Asphaug 2010).

and solving for velocity. Impact velocities at the mutual escape speed result in the accretion of the impactor and target (e.g., Safronov 1964). During planet formation however, $v_{coll} > v_{esc}$ is commonplace. This is due to the scattering of planetary bodies into more eccentric (more elliptical) or inclined (tilted with respect to the paths of other bodies) orbits by gravitational forces, increasing relative velocities. Colliding bodies thus are not generally in free fall ($v_\infty = 0$) but have an additional contribution to their relative velocity ($v_\infty > 0$). This yields a collision velocity, $v_{coll} = \sqrt{v_{esc}^2 + v_\infty^2}$, that often exceeds $v_{esc}$. Material may be ejected as a result, leading to erosion of the colliding bodies (e.g., Safronov 1972).

## 1.2. Diversity of Giant-Impact Outcomes

Early computer simulations of the giant-impact process led to the characterization of the outcomes of collisions that occur beyond $v_{esc}$ and provided new potential pathways for the formation of the Moon and Mercury. These studies also revealed that impact geometry plays a central role in dictating the fate of any colliding pair [see the sidebar titled What Is the Most Probable Collision Angle ($\theta_{coll}$)?].

**1.2.1. Moon formation in off-axis collisions.** From calculations of planet formation near 1 astronomical unit (AU), Hartmann & Davis (1975) showed that giant impacts into proto-Earth at collision velocities just above $v_{esc}$ would have been commonplace. They proposed that the Moon coagulated from a silicate-rich disk around Earth generated by a large-scale collision, providing an explanation for the paucity of volatile elements in Apollo samples. The colliding body was posited to be roughly the mass of Mars—a truly giant impact. Furthermore, radiometric age dating of Apollo lunar samples indicates that the oldest samples formed ~100 million years after the formation of the Solar System (Tatsumoto & Rosholt 1970). This was broadly consistent with giant impacts occurring after the nebular gas dissipated (Safronov 1964), ~2–3 million years into Solar System formation (e.g., Williams & Cieza 2011). Accretion of the Moon from silicate materials also provides a natural explanation for its small core, which accounts for about 10% of its mass (Asphaug 2014).

The angular momentum of the Earth-Moon system has also served as a critical benchmark for constraining the preimpact parameters of the Moon-forming impact. To form the proto-lunar disk, debris must be ejected into bound and stable orbits (Cameron & Ward 1976). This is made possible by collisions with sufficient angular momentum, which are predominantly dependent on collision angle and velocity (neglecting preimpact rotation of the two bodies). At contact, the specific angular momentum of the relative orbit is proportional to $(R_T + R_I)v_{coll} \sin\theta_{coll}$, where $\theta_{coll}$ is the collision angle (**Figure 2a**). As such, simulations of the Moon-forming giant impact involved off-axis collision geometries as a means of generating bound postimpact disks around Earth ($\theta_{coll} \gg 0°$) (e.g., Benz et al. 1986, Stevenson 1987, Cameron & Benz 1991, Cameron 2000).

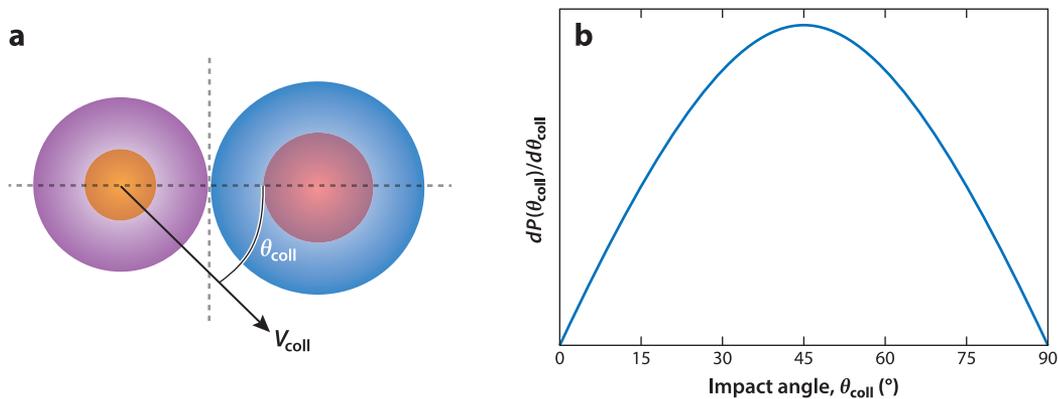

**Figure 2**

(*a*) A diagram of the geometry of a giant impact. (*b*) The differential probability distribution of the collision angle, $dP(\theta_{coll}) = \sin(2\theta_{coll})$ $d\theta_{coll}$ (Shoemaker 1962). Due to the $\sin(2\theta_{coll})$ distribution, head-on ($\theta_{coll} \sim 0°$) or glancing collisions ($\theta_{coll} \sim 90°$) are much less likely than moderately glancing geometries ($\theta_{coll} \sim 45°$).

By examining the intermediate epochs of lengthy Moon formation calculations, it was discovered that the impactor can graze the target, proceed along a gravitationally bound orbit, and eventually be accreted onto the proto-Earth in a spiraling fashion (see **Figure 3** and the **Supplemental Videos**). This graze-and-merge process liberates debris that form the circumplanetary disk from which the Moon can accrete and can occur even in relatively gentle scenarios ($v_{coll} < 1.1v_{esc}$). Refinements of the graze-and-merge formation of the Moon with a Mars-sized impactor would result in the canonical model of Moon formation (Canup 2004). In alternative, high-angular momentum, high-velocity giant-impact models (e.g., Ćuk & Stewart 2012), the postimpact Earth is a distended vapor-rich structure from which the Moon forms, a formation pathway that has become the focus of recent work (Lock et al. 2018).

### 1.2.2. Mercury formation in erosive collisions.

Just as the Moon's small core has been attributed to a giant-impact origin, Mercury's large core has been attributed to a giant impact of a particularly destructive nature (Smith 1979). As indicated by data from NASA's MESSENGER mission and by earlier measurements of Mercury's large uncompressed density, the planet's core-mass fraction of ~70% far exceeds that of the other terrestrial planets (25–35%) and the iron fraction expected from early Solar System materials (Palme & O'Neill 2003). Giant-impact simulations demonstrated that an impactor colliding at several multiples of $v_{esc}$ into proto-Mercury (a relatively high-energy collision) is indeed capable of stripping the target of most of its mantle, forming a planet with a large core (Benz et al. 1988). Ejected materials, however, may reaccumulate onto the target, counteracting the initial mantle stripping (e.g., Benz et al. 2007). As such, the giant-impact models for Mercury formation continue to undergo testing and revision (Ebel & Stewart 2018), and models of nonimpact origin continue to provide alternatives to this hypothesis (e.g., Johansen & Dorn 2022).

### 1.2.3. Hit-and-run collisions.

A central challenge to forming planets with large cores by eroding the target's mantle materials in a near–head-on giant impact is that these scenarios are generally not common (**Figure 2b**). When $v_{coll}/v_{esc}$ is high enough, an off-axis collision can result in the impactor glancing the target and escaping along deflected orbits as a runner (Agnor & Asphaug 2004). In this hit-and-run outcome, the impactor avoids accretion and continues along a deflected orbit around the central star as a body distinct from the target (Asphaug et al. 2006). To achieve the

**Canonical model of Moon formation:**
accretion of the Moon from a circum-Earth disk generated by a grazing impact between proto-Earth and Theia (a roughly Mars-sized impactor)

**Core-mass fraction:**
the ratio between the mass of the core of a planetary body and its total mass; a simplified compositional parameter

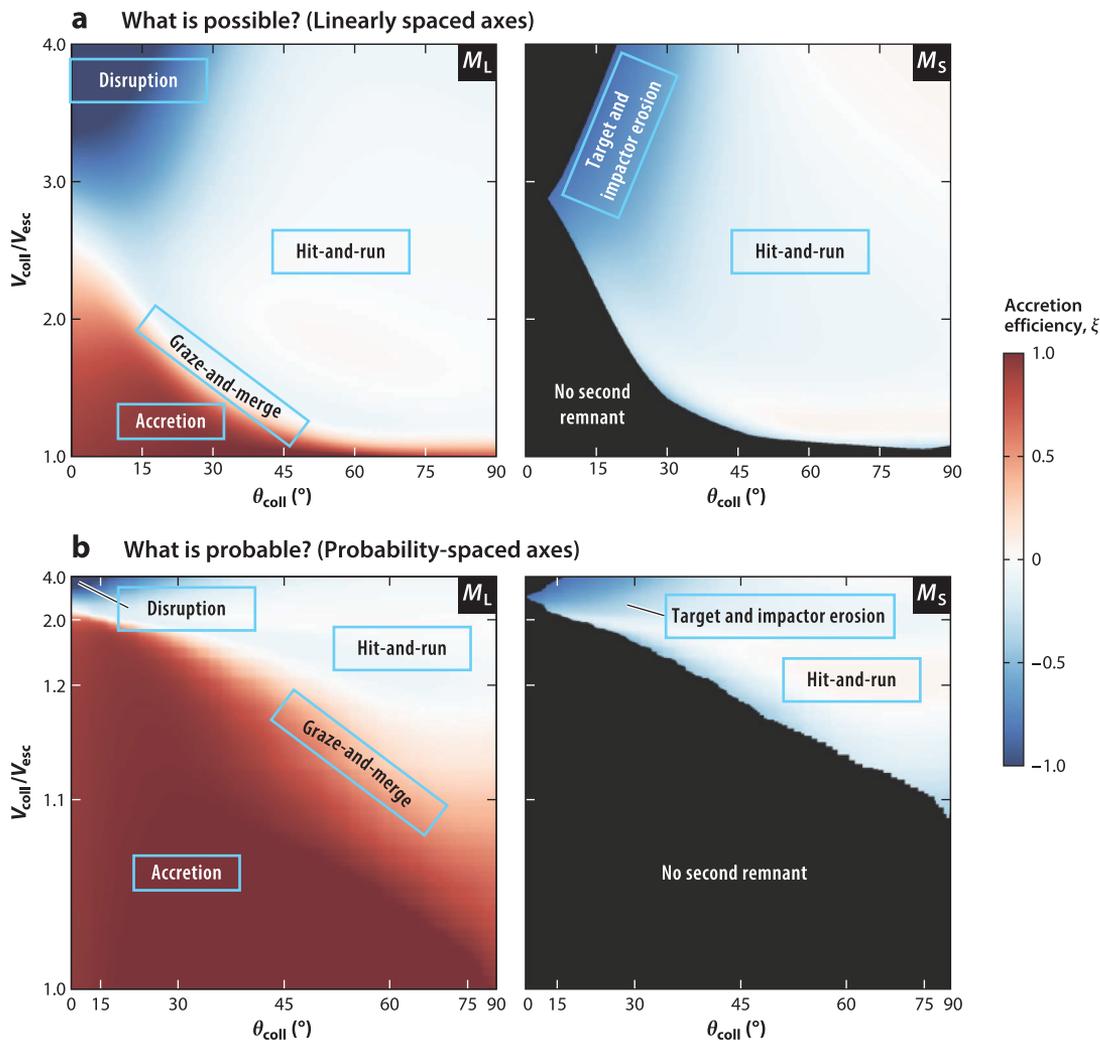

**Figure 3**

(*a*) Accretion efficiency $\xi_L = (M_L - M_T)/M_I$ and $\xi_S = (M_S - M_I)/M_I$ for the largest and second largest remnants mass $M_L$ (*left*) and $M_S$ (*right*) of a giant impact as a function of collision angle and velocity. Warm and cool colors represent mass gain and loss as a result of the collision, respectively. The target is Mars sized, and the impactor-to-target mass ratio is 0.7. Both have a core-mass fraction of 30%. (*b*) The same data as panel *a*, but axes ticks are spaced according to the expected distribution of collision angles governed by sin ($2\theta_{coll}$) and impact velocities derived by Gabriel et al. (2020) from Solar System formation simulations by Chambers (2013). Outcomes that are more probable encompass more area in panel *b*. The figure was generated using a machine-learning giant-impact model (Cambioni et al. 2019, Emsenhuber et al. 2020) and is analogous to Stewart & Leinhardt (2012), figure 5. Giant-impact simulation videos are provided in **Supplemental Videos 1–5** and can also be accessed by clicking the hyperlinks in the figure in the online PDF.

**Supplemental Material ›**

necessary velocity, the impactor must approach the target along an eccentric or inclined orbit in order to have sufficient $v_\infty$ (Jackson et al. 2018). Such a collision has high angular momentum, allowing for the production of postimpact disks from which satellites can form, motivating the exploration of hit-and-run collisions as a potential model for Moon formation (e.g., Reufer et al. 2012).

Hit-and-run collisions preferentially strip outer mantle materials of the impactor (**Figure 3**), providing a second pathway for increasing planetary core-mass fractions. The impactor is preferentially eroded because, due to its smaller mass, it receives a higher energy per unit mass in the collision and has a lower gravitational binding energy than the target. The gravitational binding energy of a planet acts to hold the planet together from disruption and is defined as

$$U_{\mathcal{G}} = \mathcal{G} \int_0^R M(r) m(r) \frac{dr}{r},$$

4.

where $m(r)$ is the mass of a shell of size $dr$, $M(r)$ is the planetary mass interior to the shell, and $R$ is the radius of the planet. Assuming constant-density bodies, the binding energy is

$$3\mathcal{G}M^2/5R,$$

5.

which is larger for the target due to its larger mass. As a result, the target tends to preferentially accrete the mantle of the runner, acting as a gravitational sink for impact debris. The effectiveness at which impactors are eroded in hit-and-run collisions and the sequestration of mantle debris by the target led to the proposal that Mercury could be the runner of a hit-and-run collision instead of the target in an erosive collision (Asphaug & Reufer 2014). Analogous to what the target experiences in an erosive head-on collision, runners in hit-and-run collisions experience hydrostatic unloading of deep mantle rocks by upward of several tens of percent of pressure, potentially generating melt and the release of volatiles (Asphaug et al. 2006).

But what is the orbital-dynamical fate of the runner in a hit-and-run collision? Planet formation studies that take into account hit-and-run outcomes have demonstrated that the runner can recollide with the target (Chambers 2013). Reimpact is likely because the deflected postimpact orbits of the two bodies intersect at the original impact point in inertial space. The returning impactor can hit-and-run as well, starting a series of interconnected recollision events called collision chains (Emsenhuber & Asphaug 2019a). However, the recollision is not immediate and has been shown to occur tens of millions of years later for chains in the Solar System near 1 AU (e.g., Emsenhuber & Asphaug 2019b). Collision chains have recently been shown to be a viable mechanism of Moon formation (e.g., Asphaug et al. 2021) because each recollision would contribute to the proto-lunar disk from which the Moon is thought to have formed. The collisions lead to homogenization between the disk and Earth, helping to explain the remarkable isotopic similarity between Earth and the Moon. The consequences of collision chains on planetary interior processes (e.g., planetary dynamo generation and mantle and core convection) have yet to be explored.

## 2. MODELING GIANT IMPACTS

Impossible to be fully replicated in physical laboratories, giant impacts are understood predominantly through computer simulations. This places important emphasis on the number of variables and processes involved in collisions, as well as the methodologies used to simulate them (see the sidebar titled The Combinatorics of Giant Impacts).

### 2.1. Dominant Variables

#### 2.1.1. Relative size of the colliding bodies.
As the sizes of the impactors relative to the targets grow, so do the consequences of their collisions. When the impactor-to-target mass ratio is much less than 1%, a crater is generated on the surface. At a mass ratio near 1%, impactors leave near-global scars. For example, the South Pole–Aitken basin covers nearly one-third of the Moon, yet its proposed impactor is estimated to be at most ∼1% of the Moon's mass (Schultz & Crawford 2011, Wieczorek et al. 2012, Melosh et al. 2017). Even this not-so-giant impact is thought to have excavated deeply, reaching the ancient lunar mantle, making the crater an important target of future exploration (e.g., Natl. Acad. Sci. Eng. Med. 2022). In giant impacts, the impactor is

## THE COMBINATORICS OF GIANT IMPACTS

The vastness of the parameter space of giant impacts becomes evident even when considering only a few of the preimpact conditions. Sampling 10 values of collision angle, impact velocity, and relative size requires $N = 10^3 = 1,000$ computer simulations, discussed later. Accounting for 10 values of core-mass fraction for the target and impactor corresponds to $N = 10^5 = 100,000$ scenarios. Accounting for the rotational angular momentum of the colliding bodies adds 6 additional parameters: 2 for the rotational rates and 4 angles to describe the angles between the rotational axes and the impact plane. Even ignoring the presence of atmospheres, third-body effects, and thermal states, a full parameter sampling requires $N = 10^{10} = 100,000,000,000$ giant-impact simulations! Far too intensive to explore in this way, giant impacts are a ripe area for the application of machine-driven methods and advancements in simulation methodology.

larger than 1% of the mass of the target. A prominent example is the Borealis basin on Mars, for which impact hypotheses demand an impactor that was 9–43% the mass of Mars to produce a hemispherical magma ocean from which the Mars northern lowlands can form (Marinova et al. 2011).

**2.1.2. Absolute size of the colliding bodies.** Material strength and gravity moderate the outcomes of collisions depending on the absolute size of the colliding bodies (Holsapple 1994). The gravitational binding energy of a body increases with its size (Equation 4). In contrast, material strength is predicted to decrease with size as larger bodies feature larger and thus weaker flaws [e.g., the so-called size-dependent strength theory (Weibull 1939)]. Flaws act as nucleation sites for fractures that lead to the failure of the material. As such, there exists a poorly understood threshold size (roughly a few hundred meters in diameter) where bodies are the easiest to erode for a given specific impact kinetic energy (e.g., Benz & Asphaug 1999). Below and above the threshold size, the impact outcomes are said to be strength dominated and gravity dominated, respectively.

Importantly, material strength can still play a role in giant impacts between large bodies that would otherwise be understood to be gravity dominated based on their size. Rheological effects such as friction and the crushing of preexisting spaces (pores) during collisions, for example, have been demonstrated to serve as a sink for impact kinetic energy, reducing the amount of ejected materials in giant impacts for bodies as large as 100 km (Jutzi 2015). The overall consequence of material strength on the collisional evolution of Solar System bodies larger than 100 km is the object of current studies (Jutzi et al. 2019, Jutzi & Michel 2020, Rozehnal et al. 2022). Even giant-impact models of Mars-scale bodies show that frictional heating causes a broader melt region in the martian mantle than when friction is neglected, which has important implications for the hypothesized Borealis impact basin (Emsenhuber et al. 2018). Thus, even where the broad outcomes of giant impacts, such as the mass of the largest remnant, are gravity dominated, other metrics such as heat and melt generation may depend critically on the absolute scale of the collision and the assumed rheology.

The presence of a core and the effects of compression under self-gravity also have important consequences for the susceptibility of planetary bodies to disruption. Under self-gravity, mantle materials become compressed at depth, causing some minerals to transition to their higher-density crystal structures, increasing the bulk density of a planet. In addition, during differentiation, higher density materials become centrally concentrated into planetary centers, forming cores. In both processes, the gravitational binding energy of the planet (Equation 4) increases, exceeding that of the constant density assumption (Equation 5). The total binding energy of the colliding system, which is the sum of the binding energies of the target and impactors (Equation 4) and the orbital

**Strength:** the ability of a material to withstand an applied load without deformation

**Friction:** a force that resists motion between solid materials as they undergo deformation

**Rheology:** the study of how rocks deform considering their physical properties and conditions when subjected to loads

**Differentiation:** process where molten planets partition denser materials into the center, generally forming a metal-rich core potential energy at contact (Equation 3)

$$U = U_{\mathcal{G},\mathrm{T}} + U_{\mathcal{G},\mathrm{I}} + U_{\mathcal{G},\mathrm{contact}}, \qquad\qquad 6.$$

can exceed the constant-density approximation by 10% or more depending on the core-mass fractions of the colliding bodies and their absolute size (Gabriel et al. 2020). This deviation is noteworthy, as the ratio of kinetic energy to the total potential ($K/U$) is a quantity used to estimate the disruption threshold for colliding planets. This is because collisions with equal values of $K/U$ result in similar outcomes across decades of the total colliding mass (e.g., Movshovitz et al. 2016). Whether this finding holds true for super-Earths and more exotic planetary compositions [e.g., dense, likely iron-rich planets in the exoplanetary catalog (Adibekyan et al. 2021)] is not currently understood.

Planetary density may also influence the transition to hit-and-run collisions. Differentiated and centrally compressed planets feature a concentration of higher-density materials and mass in the interior (e.g., planetary cores). This concentration of mass in the core translates to less material being involved in collisions at glancing angles (see the cores in **Figure 2** and Asphaug 2010, equation 9). The result provides geometrical context for recent hydrocode simulations that suggest that differentiated planets more readily undergo hit-and-run (Gabriel et al. 2020). At the same impact velocities, early planets that have yet to undergo differentiation would be more susceptible to impactor disruption (Leinhardt & Stewart 2012). And since hit-and-run collisions lengthen the timescales of planetary growth (e.g., Chambers 2013), the size of a planet's core may influence how it evolves through subsequent collisions.

### 2.1.3. Impact strain.
Impact disruption and the size distribution of debris in strength-dominated collisions are controlled in large part by material fragmentation and failure. However, there are several challenges to account for these rheological processes in giant-impact simulations. The resistance of a material to deformation (strain), fragmentation, and failure is dependent on rate of deformation. For impacts, strain rates scale approximately by $v_{\mathrm{coll}}/(2R_{\mathrm{I}})$. Models of dynamic fracturing and failure developed in the laboratory are used to make predictions at the giant-impact scale. However, due in part to the difference in scale $R_{\mathrm{I}}$, strain rates in giant impacts are several orders of magnitude lower than dynamic experiments conducted in the laboratory (e.g., Grady et al. 1977). Lab-based models are thus extrapolated to lower strain rates with significant deviations in results, motivating the development of new models based on first principles and multi-scale approaches in order to take into account failure processes in giant impacts (e.g., Ramesh et al. 2015).

Strain rate also varies depending on the proximity to the impact site, making different rheologic models relevant for various regions in the impacting bodies. The concentration of stresses closer to the interface between colliding bodies is expected to enhance the degree of material fragmentation resulting in smaller debris sizes (e.g., Grady et al. 1977, Grady & Kipp 1985). Elsewhere strain rates are lower because stresses are driven by tidal effects over gravitational timescales [minutes to hours for typical planetary densities (Asphaug 2010)]. Here, models of dynamic fracturing become relevant due to the concentration of shear stresses (Ramesh et al. 2015), and pore collapse and friction (Jutzi 2015) may also contribute to the overall heating budget and fragmentation behavior. Understanding the influence of rheological processes that are characteristic of different zones of the collision through dedicated hydrocode simulations remains an important area of future development.

### 2.1.4. Absolute velocity and thermodynamics.
If the absolute velocity $v_{\mathrm{coll}}$ is higher than the sound speed in the material, a giant impact is a supersonic event, and understanding these events demands accurate shock equations of state. Shock waves generated by supersonic collisions lead

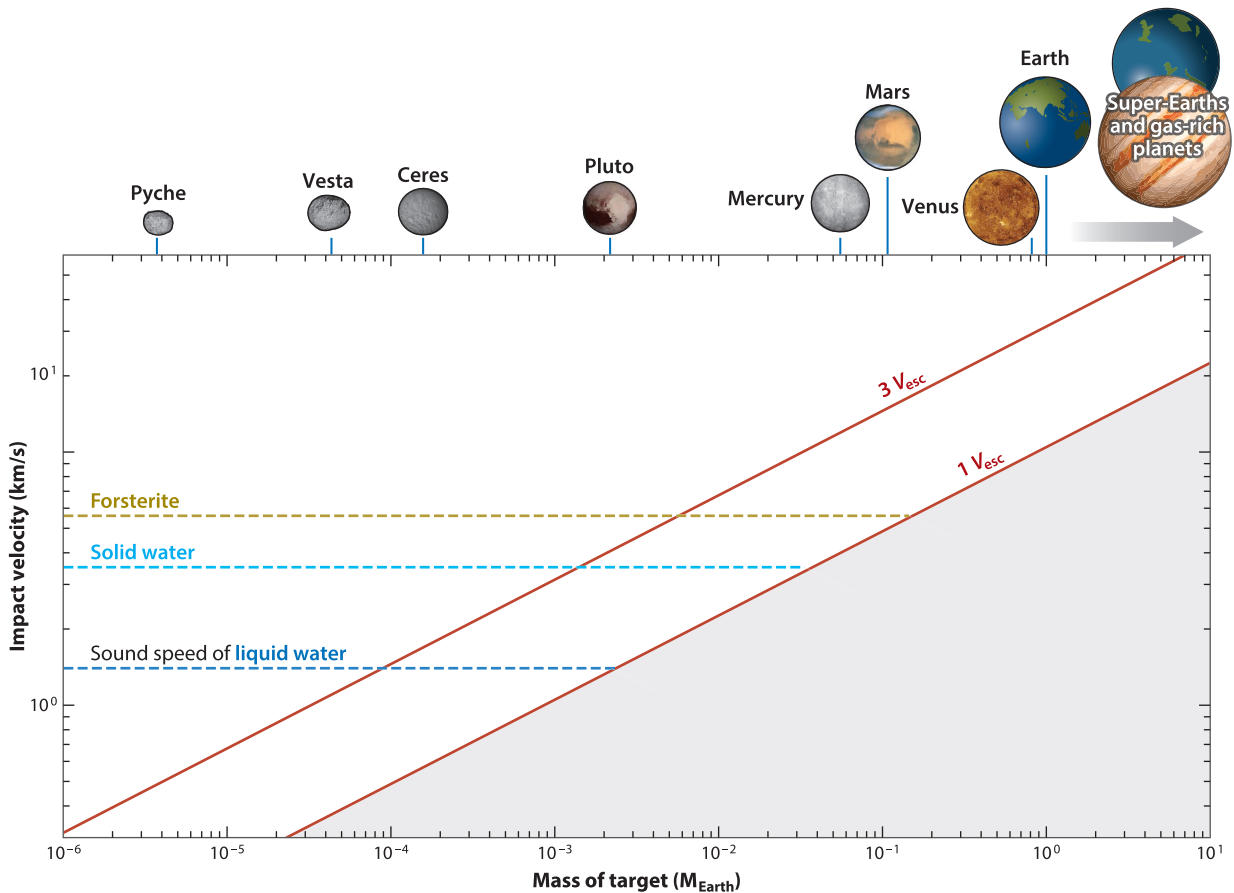

**Figure 4**

Collision velocity as a function of the mass of the target in Earth masses. The lowest possible collision velocity without the presence of nebular gas is $v_{coll} = 1\ v_{esc}$, a gentle scenario that always results in a merger of the two bodies. Solar System formation simulations report velocities $v_{coll} \approx 1–1.6\ v_{esc}$, with exceedingly few near $3\ v_{esc}$ (e.g., Agnor et al. 1999). For simplicity, sound speeds are shown for ~0°C and ~1 atm conditions, a constant density of 5 g/cm³ is assumed for the escape velocity calculation, and the impactor-to-target mass ratio is 0.7. The sizes of the planet images are not to scale.

to a discontinuous change in pressure and density across the wave front, transforming kinetic energy into heat, leading to phase change and the production of vast amounts of melt, vapor, and plasma. This effectively overwhelms the rheological processes discussed previously. Melt and vapor production from shocks are expected to be enhanced in warm planetary embryos such as those emerging from the nebular phase (e.g., Weiss & Elkins-Tanton 2013) because they are closer to thermodynamic thresholds for melting and vaporization of planetary materials (Davies et al. 2020).

**Equation of state:** a thermodynamic tool that relates variables such as pressure, volume, and density for a material

Due to the dependence of impact velocity on $v_{esc}$, the sound speed in solid water (ice) allows shock-inducing collisions to occur in giant impacts between bodies roughly the size of Pluto (**Figure 4**). The large pressures in planetary interiors and in the shock front generate high-pressure crystalline phases of water and, upon release, can produce ample steam in the process, motivating new shock equations of state to describe water under a wider range of conditions (Stewart et al. 2008). A central challenge in modeling the fate of debris in collisions involving hydrated minerals is that most equations of state implementations do not explicitly account for the devolatilization and rehydration process that would occur within impact plumes. Further developments in this area are key to refining models of icy body collisions, such as those posited for the formation of the Pluto-Charon system and the Haumea collisional family (e.g., Brown et al. 2007, Leinhardt et al. 2010, Canup 2011, Kenyon & Bromley 2014).

For rocky bodies, the dynamical evolution of giant-impact debris is likely influenced by its thermodynamic state. Debris-debris collisions between solid particles have a strong control on the lifetime and size distribution of the debris cloud (e.g., Jackson et al. 2014, Kobayashi et al. 2019). As colliding bodies grow larger than the Moon, their collision velocities are supersonic, transitioning them from producing mostly solid debris driven by material fragmentation to predominantly producing debris in the form of melt and vapor (Gabriel & Allen-Sutter 2021) (**Figure 4**). The process generates melt in planetary mantles (e.g., Nakajima et al. 2021), and in Earth-scale giant impacts ($v_{esc} > 10$ km/s) collisions cause rocks to reach stellar temperatures (Canup 2004) (see also videos in the **Supplemental Materials**). Modeling these extreme conditions at high fidelity demands continued development of new shock equations of state (e.g., Root et al. 2018) and the inclusion of thermal radiation, chemical disequilibrium, and other complex processes in calculations.

### 2.1.5. Preimpact rotation.
The relative spin rate and orientation of colliding bodies are important yet poorly understood factors influencing impact outcomes. For example, preimpact rotation may enhance the amount of material released to a postimpact disk. In extreme conditions, a body rotating near the limit of its stability requires only a small perturbation to deliver mass shedding consequences (e.g., Dobrovolskis 1990). This mechanism of angular momentum transfer from the rotating planets to postimpact disks naturally motivates the exploration of preimpact rotation on the formation of the Moon (e.g., Rufu et al. 2017). In addition to potentially contributing to the generation of disks, preimpact rotation can eject materials into different orbits than expected from models of collisions between nonrotating bodies. Thoroughly parameterizing the effects of preimpact rotation on giant impacts, however, continues to present a major challenge, as it entails expanding the parameter space considerably beyond the few basic parameters of mass ratio, velocity, and collision angle.

## 2.2. Computer Simulations

The outcomes of giant impacts are predominantly studied through computer simulations and in particular the Smoothed-Particle Hydrodynamics (SPH) method (**Figure 5**). The advancement of computational power has allowed for the wider exploration of preimpact conditions, such as core-mass fraction and material type (Marcus et al. 2009, 2010; Reufer 2011). Improvements in resolution have also allowed for postimpact disks and the debris-size distribution to be understood in greater detail (e.g., Nakajima & Stevenson 2014, Genda et al. 2015). Furthermore, novel machine-learning methods now enable generalizing the results of sparse data sets of SPH simulations to provide predictions over the relevant parameter space (e.g., Cambioni et al. 2019, Timpe et al. 2020).

Developed in the 1970s for stellar astrophysics applications (e.g., Gingold & Monaghan 1977), such as the mergers of stars and galaxies, SPH was extended to giant-impact studies in the 1990s (e.g., Benz & Asphaug 1999) and continues to undergo refinements today (e.g., Monaghan 2012). The method discretizes the model domain (planets) into mass particles (sometimes referred to as nodes), ensuring conservation of mass throughout the simulation. Particles move according to fluid motion described by the Navier-Stokes hydrodynamic equations of motion (for a review, see Violeau 2012). Each particle serves as a point of interpolation for continuous fluid properties

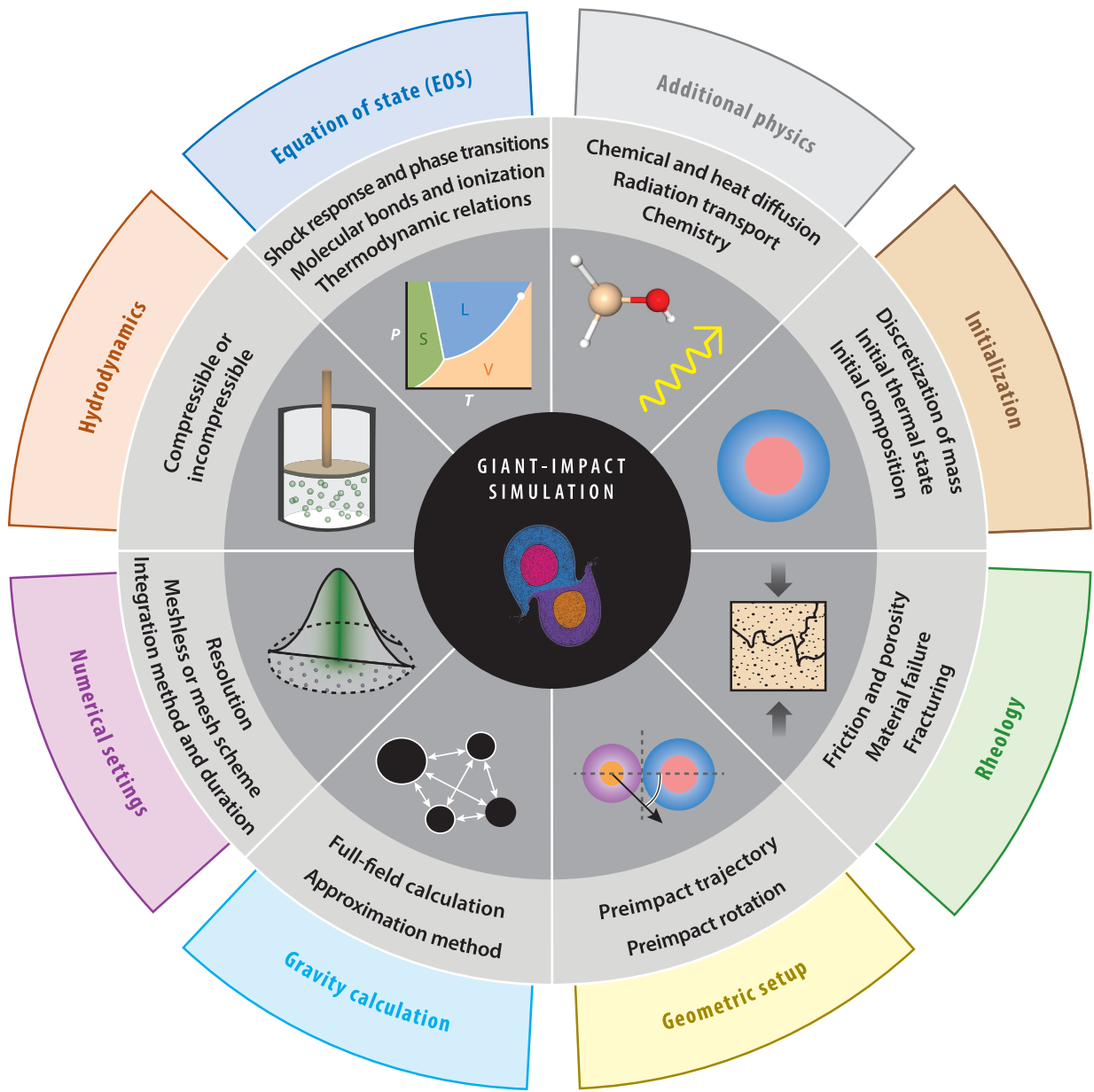

**Figure 5**

The physics and computational modules and approaches involved in performing computer simulations of giant impacts. The inclusion of additional physics modules, such as radiation transfer and chemical reactions, lies at the forefront of giant-impact modeling.

according to a smoothing kernel that defines a variable's value in space, and the kernel function can vary depending on the SPH setup. In regions where no particles are present, the equations of motion are not solved. This is inherently beneficial to calculations that involve material transport across long distances in the vacuum of space typical of giant impacts, saving computational resources. Shock discontinuities in SPH are commonly smeared across several particles using artificial viscosity, which ensures that momentum fields are differentiable and provides a shock heating mechanism (e.g., Gingold & Monaghan 1977). The pairwise force of gravity between particles would normally require computationally expensive calculations of order $N^2$ (where $N$ is the number of SPH particles). Instead, planetary and astrophysical SPH codes implement approximation methods, such as hierarchical tree-based algorithms, which afford critical improvements to the rapid calculation of gravity (Barnes & Hut 1986), enabling higher-resolution simulations for a given run time.

Similar to other methods, accurate interpretation of SPH results requires taking into account possible artifacts that depend on simulation design. The number of SPH particles over which the planets are subdivided defines the simulation resolution and has a marked effect on outcome. The highest-resolution simulations now approach tens of millions of particles, but a planet divided into millions of equal-mass particles still has only hundreds along the radial dimension, limiting the resolution of smaller-scale flows, turbulence, and the debris field. This shortcoming produces simulation outcomes, such as the mass of the largest remnant, that vary as a function of resolution (e.g., Genda et al. 2015). Generally, greater particle resolution reduces the overall influence of artificial tension forces between particles that can arise in SPH (Agertz et al. 2007). Importantly, simulation outcomes tend to converge to a particular value as resolution increases. It is thus critical to perform what are called resolution convergence tests, where resolution is increased incrementally until the convergence of an output of interest can be inferred. For common choices in resolution, the impact energy necessary to disrupt two planets by a giant impact may be underestimated by tens of percent (Genda et al. 2015). Reaching resolution convergence in the angular momentum of the postimpact disk—a key parameter to benchmark models of Moon formation—also remains challenging, even at very high particle counts (Hosono et al. 2017). Obtaining higher resolution in general enables higher-order flow in simulations (more accurate material mixing), which is important in developing accurate chemical equilibration models in planetary evolution studies.

Finally, the stability of vapor-rich postimpact structures common to large-scale collisions (e.g., Lock & Stewart 2019) is intimately influenced by SPH's sensitivity to artificial viscosity. Erroneous triggering of artificial viscosity and momentum transport in shear flows (e.g., Okamoto et al. 2003) allows for the unphysical collapse of pressure-supported disks (Raskin & Owen 2016). Depending on the form of artificial viscosity used in an SPH simulation, the production of impact heat, vapor, and melt can vary (Gabriel & Allen-Sutter 2021). These various artifacts are the topic of current important efforts to improve the SPH method (e.g., Read et al. 2010, Monaghan 2012, Frontiere et al. 2017) and represent important hurdles for giant-impact science.

## 3. GIANT IMPACTS IN TERRESTRIAL PLANET FORMATION

Giant impacts are expected to occur throughout planet formation, under different conditions depending on the presence or absence of nebular gas and the mass distribution of the bodies. For terrestrial planets, however, signatures of earlier planet formation processes can be overwritten by the giant impacts that follow. New models of giant-impact outcomes and the application of probability theory provide pathways for disentangling the complex nature of giant-impact histories.

### 3.1. The Aftermath of Early Accretion

The formation and properties of planetary embryos set the initial conditions of the giant-impact phase. These initial conditions, which remain debated, influence the nature and number of giant impacts, thus producing different impact histories for the forming planets.

### 3.1.1. Giant impacts in traditional formation theory.

The traditional view of planet formation is that terrestrial planets form through collisions among planetary embryos (e.g., Wetherill 1976). Before nebular gas dissipation, mass is distributed within a large number of planetesimals with low relative velocities. As a few larger bodies coalesce, they undergo oligarchic growth by preferentially accreting the smaller bodies. This process is expected to create lunar-to-Mars-mass planetary embryos over $\sim 10^5$ years in the Solar System (Lissauer 1993, Kokubo & Ida 1996). After the nebular gas dissipates, gravitational interactions between the largest bodies lead to crossing orbits. Chaotic growth ensues, generating giant impacts occurring across tens to hundreds of millions of years (Lissauer 1993, Kokubo & Ida 1996, Chambers 2013).

### 3.1.2. Giant impacts in pebble accretion theory.

The recent theory of pebble accretion proposes that the drag by nebular gas enhances the accretion of millimeter- to centimeter-sized particles (referred to as pebbles) onto planetesimals, forming protoplanets within a few million years (e.g., Levison et al. 2010, 2015; Johansen & Lambrechts 2017; Lambrechts et al. 2019). When applied to the formation of terrestrial bodies in the Solar System, pebble accretion models report a decrease in the number of embryos by a factor of $\sim 2$ compared to classical Solar System formation theory. This corresponds to an order of magnitude fewer giant impacts (e.g., Voelkel et al. 2021) or just a single, Moon-forming giant impact (Johansen et al. 2021), placing critical emphasis on the nature of that singular collision and its detailed outcome.

## 3.2. Giant Impacts and Planetary Diversity

A simple statistical exercise can provide insight into the effect of the nature and number of giant impacts on planetary diversity. Assuming that accretion and hit-and-run are equally probable outcomes during the late stage of terrestrial planet formation, Asphaug & Reufer (2014) showed that one can evaluate impact outcomes of $N$ embryos picked out of a jar assuming either accretion (i.e., one of the bodies is removed from the jar) or hit-and-run (i.e., both bodies survive and are put back in the jar). They found that each final planet in their model survives two hit-and-runs on average. Every time an impactor survives as a runner, it loses part of its outermost silicate mantle. The greater number of times this occurs in this demonstration, the greater amount of mantle erosion the body experiences, increasing its core-mass fraction. As a result, the onset of chains of hit-and-run collisions is expected to increase the diversity of planetary composition.

To track the diversity of the population, we repeat the exercise of Asphaug & Reufer (2014) and measure the planetary diversity of the evolving embryos as the number of different hit-and-run species over the total population size (i.e., its statistical richness) (**Figure 6**). A hit-and-run species encompasses those embryos that have survived the same number of hit-and-run collisions and have similar core-mass fractions. We assume that the initial population is composed of embryos with the same core-mass fraction (that is, diversity of 0%) and measure how diversity of core-mass fraction increases as a function of number of giant impacts. If long chains of giant impacts occur, planetary diversity increases rapidly. If instead just a handful of giant impacts occur, as expected in some pebble accretion models for planet formation, a lower degree of planetary diversity is expected.

Depending on the number of collisions, the giant-impact process can be so transformative for the embryos that the information about the nature of the primordial population can be partially or completely lost. The primordial population is transformed through giant impacts due to the coupled effect of hit-and-run collisions, which increase the diversity of planetary compositions, and the loss of primordial bodies through accretion onto the larger ones. The latter process—defined as accretionary attrition by Asphaug & Reufer (2014)—is akin to a loss of genetic history by extinction and may induce survivorship biases in planet formation studies. However, it is noteworthy

---

**Planetesimals:**
roughly 100-km-sized bodies that form early in the planet formation process in the presence of nebular gas

**Survivorship bias:**
the error of inferring conclusions from outcomes that made it past a selection process and overlooking those that did not

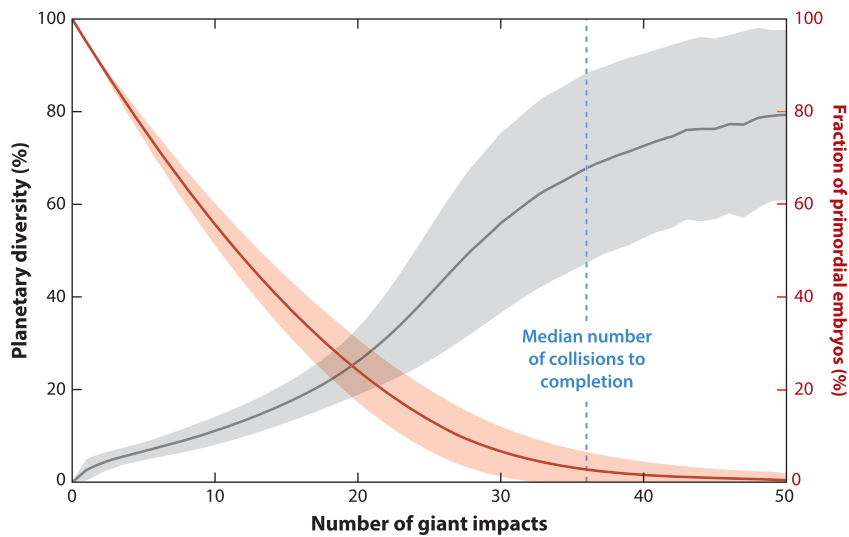

**Figure 6**

Left axis: planetary diversity as a function of the number of giant impacts. Diversity is measured as the number of different hit-and-run species—that is, the subset of the population whose bodies have survived the same number of hit-and-run collisions. Right axis: the fraction of bodies equal to those of the initial population, whose bodies are identical to one another and never participated in a hit-and-run collision. The solid curves are the average value of 10,000 runs, assuming that accretion and hit-and-run have 50–50% probability of occurring as in Asphaug & Reufer (2014). Shaded areas indicate the corresponding 1-standard-deviation bands. The results do not change significantly when the hit-and-run occurrence rate is varied by $\pm10\%$ [the variability observed in studies in the literature (e.g., Chambers 2013, Emsenhuber et al. 2020)].

that, in the simple exercise of **Figure 6**, accretion alone is not sufficient to induce the complete loss of information. Indeed, if we assume that all collisions lead to perfect accretion (another very unlikely scenario), the embryos would become larger versions of their previous selves but new hit-and-run species would not form, unrealistically keeping planetary diversity of core-mass fractions to 0%. This is potentially at odds with the high core-mass fractions of Mercury and the high density of some M-type asteroids (Asphaug & Reufer 2014) and some super-Earths (Adibekyan et al. 2021). Importantly, these bodies may also form in erosive high-velocity, low-impact-angle collisions, but these events are expected to be rare, at least in Solar System formation (e.g., 10–15% of the collisions) (Emsenhuber et al. 2020).

### 3.3. Modeling Giant Impacts in Planet Formation Studies

*N*-body simulations have greatly advanced our understanding of Solar and extra-Solar System formation (for reviews, see Morbidelli et al. 2012, Raymond & Morbidelli 2022). These simulations model the heliocentric orbits and accretion of growing bodies for millions of years and have confirmed the central role of giant impacts in shaping the diversity of planets (e.g., Stewart & Leinhardt 2012, Chambers 2013, Quintana et al. 2016, Scora et al. 2020). Here, we review the three predominant methods to model giant impacts in *N*-body simulations.

**3.3.1. Brute-force approach.** The most rigorous and computationally demanding method is to hand off each collision to an SPH simulation for explicit modeling, pause the *N*-body integration temporarily, and then feed the results back into the planet formation simulation (e.g., Genda et al. 2017, Burger et al. 2020). The challenge, however, is that each SPH simulation can require days

## PERFECT MERGING AND OTHER OVERSIMPLIFICATIONS IN COLLISION MODELING

Before the advent of fast computers and giant-impact models, two colliding bodies of mass $M_T$ and $M_I$ were often assumed to perfectly merge into a body of mass $M_T + M_I$ without losses to debris. However, decades of computer simulations have shown that this is an oversimplification (e.g., Asphaug et al. 2006, Leinhardt & Stewart 2012). Furthermore, as the masses perfectly merge, so do the rotational angular momenta (Agnor & Asphaug 2004), which would lead to unphysically high spin rates (Agnor et al. 1999). To hasten $N$-body planet formation calculations, artificial inflation factors for planetary radii have often been used to achieve accretion even in noncollisional trajectories. These shortcuts artificially decrease the amount of material transport throughout the system, reduce formation timescales, and reduce the degree of dynamical stirring through the lack of close encounters (e.g., Süli 2021, Haghighipour & Maindl 2022), demonstrating the importance in accounting for inefficient accretion.

of computer time, even at modest resolution (e.g., Vacondio et al. 2021). $N$-body simulations may take months to complete, depending on the number $N$ of bodies. For hundreds of giant impacts, calculations can require a year to run (see the sidebar titled Perfect Merging and Other Oversimplifications in Collision Modeling). Thanks to recent computational improvements [e.g., $N$-body simulations using graphics-processing units (Grimm & Stadel 2014, Woo et al. 2021)], brute-force approaches may be just within reach.

### 3.3.2. Scaling-law approach.
Scaling laws are mathematical equations that help to predict the outcome of a physical process based on the results from the same process under different conditions [i.e., scaling results under certain conditions to another (Buckingham 1914)]. For example, experimental data of impacts into small targets may be used to predict the outcomes of impacts into larger targets (e.g., asteroids). In this case, target size is being scaled, but other variables such as velocity can be chosen. Traditionally, scaling laws have been fit to a range of different data types such as crater observations, nuclear/explosive tests, laboratory experiments of hypervelocity impacts, and hydrocode output (for a review, see Holsapple 1993). The challenge of using scaling laws for giant impacts is in the choice of the scaling variable. In impact cratering the point-source approximation is used to choose scaling variables based on the assumption that the deposition of energy occurs at a single location (Dienes & Walsh 1970). However, in giant impacts, bodies are similar in size, violating the point-source approximation. The need to revise scaling-law variables led to the proposal that impact disruption is equal for equal values of the scaling variable $K/U$ (e.g., Movshovitz et al. 2016). However, supersonic collisions produce deviations from this relation whose origins are uncertain (Gabriel et al. 2020). Thus, understanding the nature of rheologic and thermodynamic transitions at various scales and building unified scaling laws that span across these regimes remain important endeavors.

### 3.3.3. Machine-learning approach.
Supervised machine-learning models are parametric functions that are trained to map inputs into outputs based on a preexisting data set of examples. In the context of giant impacts, machine-learning models are parametric functions that relate preimpact parameters to impact outcomes and are trained on preexisting hydrocode simulation data sets (Cambioni et al. 2019, 2021; Valencia et al. 2019; Emsenhuber et al. 2020; Timpe et al. 2020). Most of the data set is used to tune the parametric model to fit the output data, and procedures are adopted to avoiding overfitting. The remaining data are used to assess whether the trained model can accurately predict the outcome of collision scenarios not seen during training.

**Overfitting:** when a function accurately predicts data to which it was fit but performs poorly on data to which it was not Scaling laws are based on and provide intuitions about the physics of giant impacts, while machine-learning models provide ultimate accuracy within the confines of existing simulation results and are trained without physical assumptions (for further discussion, see Cambioni et al. 2022). Caution should be paid when extrapolating the predictions of both methodologies beyond the limits of the data sets used to build them. Similar to how scaling laws can break down in certain regimes, a machine-learning model trained to predict outcomes of collisions between rocky planets should not be used for predicting the outcomes of collisions between sub-Neptunes, for example. There the initial collision is entirely between atmospheric envelopes, producing unique relationships with respect to the preimpact conditions (e.g., Marcus et al. 2010, Liu et al. 2015, Denman et al. 2022).

A rich venue of future research lies at the intersection between the two approaches. For example, machine learning can be used to identify previously unknown dependencies between impact conditions and outcomes by means of pattern recognition through unsupervised classification methods, aiding the formulation of new scaling laws. The mathematical framework of scaling laws and machine-learning models can also be used to incorporate new physical processes in planetary formation models. Typically, processes such as crust formation, geodynamical evolution, atmosphere formation, and the orbital evolution of debris are not simultaneously included in $N$-body studies due to the high computational cost of performing these calculations in parallel. This impedes our ability to match the output of $N$-body studies to planetary observations for hypothesis testing. Crustal evolution, for example, is intimately tied to the planet's impact history. Ejected debris in the canonical Moon formation model can recollide with the terrestrial planets over 100 million years, depositing an amount of debris equivalent to a kilometer-scale or deeper layer, depending on the planet (Jackson & Wyatt 2012). This occurs despite the minor amount of total debris mass in graze-and-merge compared to other collision styles. Hit-and-run and erosive collisions shown in **Figure 3** can eject even larger amounts of debris (Leinhardt & Stewart 2012). Fitting scaling laws or training machine-learning algorithms on data from high-resolution simulations of these processes and implementing them in $N$-body studies will significantly advance the realism of planet formation models, and careful tracking of how the uncertainties of these processes propagate through the entire calculation is warranted.

## 3.4. Inverting Giant-Impact Outcomes

For planetary bodies suspected to be giant-impact remnants (e.g., the Moon, Mercury), previous studies have focused on how to deterministically infer impact conditions from observations of their properties, such as mass and composition. However, a major challenge to understanding and solving this inverse problem is that the solution may not be stable. As such, a probabilistic treatment of the giant-impact process may be preferred.

To highlight the instability of the solution to the inversion problem, in **Figure 7** we use a giant-impact model to understand what impact scenario(s) may lead to a Mercury-like planet based on its mass and iron content. We explore two possible pathways: The planet is an eroded target body (**Figure 7a**, top-left panel) or an eroded impactor surviving a hit-and-run collision (**Figure 7a**, top-right panel). Even when we assume to know the masses of the colliding bodies, the solution exists, but it is not unique: Several combinations of impact velocity and angle lead to the same outcome. Including the possibility of chains of collisions considerably expands the range of potential scenarios (**Figure 7b**). Observations of additional iron-rich bodies in exoplanetary systems may add constraints on the impact scenarios required to erode mantle materials and thus help rule out regions of the solution space. However, if the number of iron-rich planets remains small, statistical assessments may not be robust (e.g., Dean & Dixon 1951). In this context, the solution

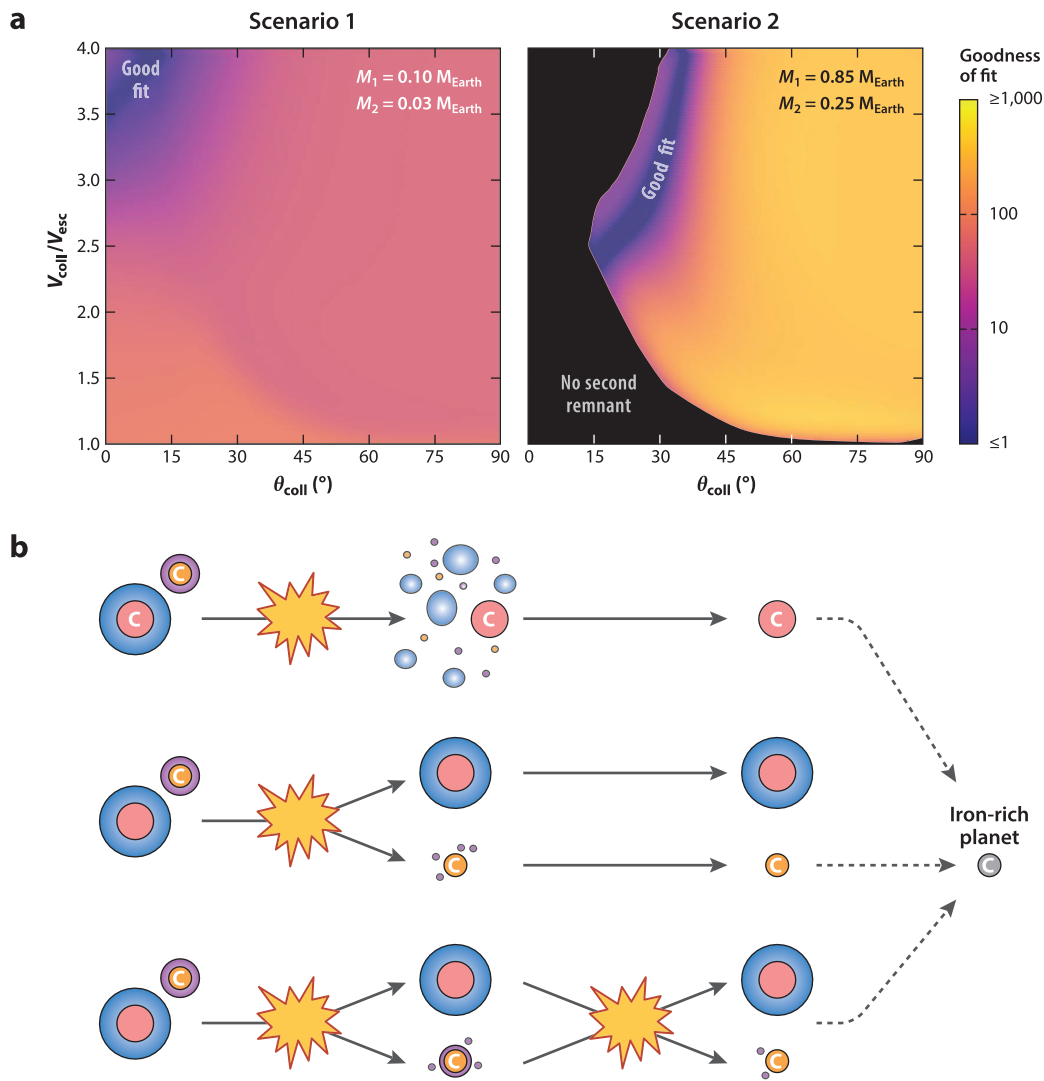

**a**

Scenario 1

Scenario 2

Goodness
of fit

$M_1 = 0.10$ M$_{Earth}$
$M_2 = 0.03$ M$_{Earth}$

$M_1 = 0.85$ M$_{Earth}$
$M_2 = 0.25$ M$_{Earth}$

Good
fit

Good fit

No second
remnant

$V_{coll}/V_{esc}$

$\theta_{coll}$ (°)

**b**

Iron-rich
planet

**Figure 7**

(*a*) Goodness of fit of predictions from two giant-impact scenarios that form a Mercury-like planet with a mass $M = 0.05$ Earth masses and a core-mass fraction of ∼70%. The planet can be either the largest remnant (Scenario 1) or a runner (Scenario 2). Mass is predicted using `collresolve` (Cambioni et al. 2019, Emsenhuber et al. 2020) and core-mass fraction using the model by Cambioni et al. (2021), assuming an initial core-mass fraction of 30%. Goodness of fit is computed as $((M - M_{obs})/\sigma_M)^2 + ((Z - Z_{obs})/\sigma_Z)^2$, where $\sigma_M$ and $\sigma_Z$ are assumed uncertainties for $M_{obs}$ and $Z_{obs}$. (*b*) Different giant-impact histories can result in an iron-rich body (indicated with the letter C): a catastrophic collision (e.g., Benz et al. 2007), a hit-and-run (e.g., Asphaug & Reufer 2014), or a chain of hit-and-run collisions (e.g., Chau et al. 2018).

may be unique but still not stable, as it can change when additional data are considered. Finally, in any hit-and-run scenario where the iron-rich planet is an eroded impactor, the question remains: What is the fate of the target?

A possible way to achieve a stable inverse problem and deal with the challenge of small number statistics is to compare the probabilities $P(x|y)$ associated with different giant-impact scenarios. This can be performed using Bayes' theorem (e.g., Cambioni et al. 2022), by treating impact conditions $x$ as probability distributions with associated prior information $\pi_{pr}(x)$ and computing the probability $P(x|y)$ that an impact scenario $x$ leads to the observed impact outcome $y$ as

$$P(x|y) = \frac{\pi_{pr}(x)P(y|x)}{\int_x \pi_{pr}(x)P(y|x)dx}, \qquad 7.$$

where the probability $P(y|x)$ of observing $y$ given the impact scenario $x$ is sampled using a giant-impact model $y = f(x)$. This typically requires computing thousands of model evaluations to determine $y$ under various conditions. Instead of computing new giant-impact hydrocode simulations for each evaluation, surrogate models (machine learning and/or scaling laws) that are fit to preexisting simulations can be used. This Bayesian framework also tackles the small number statistics challenges because it does not assume that the data are normally distributed, which is justified only for large data sets (e.g., Kruschke 2010).

The solution of Equation 7 strongly depends on the choice of the prior information $\pi_{pr}(x)$. For the collision angle, $dP \sim \sin 2\theta_{coll}d\theta_{coll}$ is the appropriate prior $\pi_{pr}(\theta_{coll})$ (**Figure 2**). Prior information for initial compositions is largely derived from measurements of meteorite composition and stellar abundance (e.g., Rubie et al. 2016, Adibekyan et al. 2021) because compositional information for most large asteroids and exoplanets is limited to their bulk density (DeMeo & Carry 2014, Dorn et al. 2015). The velocity distribution from studies of late planet formation could be used to exclude certain styles of collisions [e.g., just a few percent of high $v_{coll}/v_{esc}$ disruptions are expected (**Figure 3**)] and knowledge of the largest bodies in a swarm may help constrain the absolute scale of the expected collisions. Important caveats are that impact velocities vary in time and in location during planet formation, and velocities also vary depending on how impacts are handled in those calculations (e.g., Chambers 2013, Emsenhuber et al. 2020). The size-frequency distribution of asteroids or the formation of meteorites such as mesosiderites and pallasites may serve as important constraints on the nature of Solar Systems collisions (e.g., Scott et al. 2001, Windmill et al. 2022). Recent observations of debris fields in young exoplanetary systems that are thought to be impact remnants (e.g., Meng et al. 2014, Wyatt 2021) may also be used as benchmarks for $N$-body simulations of impact debris (Jackson & Wyatt 2012). As such, continued (exo)planetary exploration and model development, particularly for planets such as Venus, Uranus, Neptune, and others whose interior structures and compositions are poorly constrained, are essential to build reliable a priori information for determining impact conditions from present-day observations.

## 4. CONCLUSIONS

Giant impacts are global-scale, high-energy events that occur between similar-sized planets at velocities ranging from hundreds of meters to tens of kilometers per second. These events can significantly change the composition, atmosphere formation, and habitability of planets. The physics of giant impacts is rich, involving material rheology such as friction, strength, and compaction, and shock physics processes, which have important thermodynamic consequences that influence the nature and evolution of impact debris. Over several decades, computer simulations have shown that outcomes range from accretion of the impactor by the target to erosion of the target, potentially leading to distended postimpact structures dominated by vapor and melt, depending on impact velocity. Off-axis collisions are common, leading to the transient or permanent survival of the impactor depending on the velocity. Long collision chains are also possible, with unknown consequences for debris disks around the postimpact bodies.

For these reasons, giant impacts are understood to be important contributors to the observed diversity and configuration of the Solar System terrestrial planets, asteroids, and exoplanets. However, many challenges and exciting opportunities for future work remain. The transition between different collision regimes is unclear due to the poorly understood sensitivity of outcomes to parameters such as preimpact rotation; the equation of state; and the relative importance of material strength, gravity, and vaporization at different scales. The postimpact evolution of giant impacts also likely demands the addition of important, yet commonly unmodeled, physical and chemical processes. Linking giant-impact diversity to planetary diversity and pinpointing unique scenarios for the formation of suspected giant-impact remnants also remain challenging. This is due to the chaotic nature of planetary gravitational interactions and the dependence of planetary diversity on the number and types of giant impacts. The Moon-forming giant impact is the event for which we may have the most constraints; yet, many collision hypotheses remain plausible. At the frontier, improving hydrocode calculations; building a statistical understanding of giant impacts; taking advantage of the growing data set of exoplanets and debris disks; and developing new scaling laws, machine-learning methods, and *N*-body simulations are key endeavors.

## FUTURE ISSUES

1. Improving the treatment of heating and including additional physics to model postimpact effects in planetary hydrocodes are necessary steps toward testing competing impact hypotheses for the formation of the Moon, Mercury, and other suspected giant-impact products.

2. New machine-learning models and scaling laws would enable more detailed and realistic outcomes of giant impacts to be included in *N*-body planet formation studies.

3. Dedicated modeling of debris and other improvements to *N*-body and SPH simulations are necessary to interpret recent observations of debris disks around nearby stars and can provide new insights into the formation of small Solar System bodies.

4. The nonuniqueness of giant-impact inverse problems and the small number statistics of planetary accretion can be tackled through statistical frameworks such as Bayesian inference.

## DISCLOSURE STATEMENT

The authors are not aware of any affiliations, memberships, funding, or financial holdings that might be perceived as affecting the objectivity of this review.

## ACKNOWLEDGMENTS

The authors contributed equally to this review, and their order was determined by coin flip. T.S.J.G. acknowledges funding from the U.S. Geological Survey Astrogeological Science Center. S.C. acknowledges funding from the Crosby Distinguished Postdoctoral Fellowship of the Department of Earth, Atmospheric and Planetary Sciences of the Massachusetts Institute of Technology. The authors thank the reviewer, H. Allen-Sutter, E. Asphaug, S. Dibb, A. Emsenhuber, Z. Essack, Z. Lin, J. Lightholder, E. Mansbach, C. Raskin, and B. Weiss for feedback. Any use of trade, firm, or product names is for descriptive purposes only and does not imply endorsement by the U.S. Government.

# Contents

## Errata

An online log of corrections to *Annual Review of Earth and Planetary Sciences* articles
may be found at http://www.annualreviews.org/errata/earth

# Related Articles

Lipid Biogeochemistry and Modern Lipidomic Techniques
*Bethanie R. Edwards*

From the ***Annual Review of Materials Research***, Volume 52 (2022)

Biomineralized Materials for Sustainable and Durable Construction
*Danielle N. Beatty, Sarah L. Williams, and Wil V. Srubar III*

Brittle Solids: From Physics and Chemistry to Materials Applications
*Brian R. Lawn and David B. Marshall*

Material Flows and Efficiency
*Jonathan M. Cullen and Daniel R. Cooper*

Architectural Glass
*Sheldon M. Wiederhorn and David R. Clarke*